\documentclass{aa}
\usepackage{graphicx}

\def\spose#1{\hbox to 0pt{#1\hss}}
\def\lta{\mathrel{\spose{\lower 3pt\hbox{$\mathchar"218$}}
        \raise 2.0pt\hbox{$\mathchar"13C$}}}
\def\gta{\mathrel{\spose{\lower 3pt\hbox{$\mathchar"218$}}
        \raise 2.0pt\hbox{$\mathchar"13E$}}}
\def\msol{M$_\odot$}

\begin{document}

\title{VY Sculptoris stars as magnetic CVs}

\author{Jean-Marie Hameury\inst{1}
and Jean-Pierre Lasota\inst{2}}
\offprints{J.-M. Hameury}

\institute{
UMR 7550 du CNRS, Observatoire de Strasbourg, 11 rue de
           l'Universit\'e, F-67000 Strasbourg, France \\
email: {\tt hameury@astro.u-strasbg.fr}
\and
UMR 7095 du CNRS, Institut d'Astrophysique de Paris,
98bis Boulevard Arago, 75014 Paris, France\\
email : {\tt lasota@iap.fr}
}

\date{Received / Accepted}
\titlerunning{VY Scl}
\authorrunning{Hameury \& Lasota}

\abstract{We show that the absence of outbursts during low states
of VY Scl stars is easily explained if white dwarfs in these
systems are weakly magnetized ($\mu\gta 5\times 10^{30}$ G
cm$^{3}$). However, some of the VY Scl stars are observed to
have very slow declines to minimum and similarly slow rises to
maximum. The absence of outbursts during such {\sl intermediate}
(as opposed to {\sl low}) states, which last much longer than
typical disc's viscous times, can be explained only if accretion
discs are absent when their temperatures would correspond to an
unstable state. This requires magnetic fields stronger than those
explaining outburst absence during low states, since white dwarfs
in this sub-class of VY Scl stars should have magnetic moments
$\mu\gta 1.5\times 10^{33}$ G cm$^{3}$ i.e. similar to those of
Intermediate Polars. Since at maximum brightness several VY Scl
stars are SW Sex stars, this conclusion is in agreement with
recent claims about the magnetic nature of these systems.
\keywords{accretion, accretion discs -- instabilities -- (Stars:)
novae, cataclysmic variables -- (stars:) binaries : close} }

\maketitle

\section{Introduction}

Cataclysmic variables (CVs) are close binary systems in which a
white dwarf primary star accretes matter lost from a Roche-lobe
filling, low-mass secondary companion. In most CVs, matter
transferred from the secondary forms an accretion disc around the
white dwarf. Orbital periods of CVs extend from a sharp minimum at
$\sim 80$ min to $\sim 10$ hrs (a few systems with giant or
sub-giant secondaries have longer periods), with a prominent
deficit of systems between $\sim 2$ and $\sim 3$ hours (the
``period gap''). CVs are divided into different types according to
their observed properties. Some of these properties reflect the
way the systems are seen (e.g. their inclination) but most are a
reflection of the physical parameters of the system. Here we will
be mainly interested in two broad types of CVs which are
classified according to the value of the rate at which matter is
transferred from the secondary. These are the Dwarf Novae (DNs)
and Nova-Like (NLs) variables -- a nomenclature reflecting
misunderstandings of historical interest only (see Warner
\cite{w95} for details). DNs show 2-5 mag (up to 8 mag in some
cases) outbursts lasting from a couple of days up to more than a
month. Outbursts are separated by quiescent intervals lasting from
$\sim 10$ d to tens of years. It is believed that DN outbursts are
due to a thermal-viscous accretion-disc instability occurring at
temperatures where hydrogen recombines. According to the disc
instability model (DIM) which is currently used to describe DN
outburst cycle (see Lasota \cite{l01} for a review), the condition
for the instability to occur can be expressed in terms of a
critical rate at which matter is fed to the outer disc's regions.
Above this critical rate (whose value strongly increases with the
disc's radius) accretion discs are hot and stable. CVs with such
discs belong to the class of Nova-Like variable, which are defined
as `non-eruptive' CVs. There exists also a second critical
accretion rate below which a CV disc is cold and stable. The two
critical rates have the same radial dependence.

The membership of a class is not permanent but depends on the
actual mass-transfer rate. It seems, for example, that a NL should
become a DN if a fluctuation brings the mass-transfer rate below
the critical value for instability (the opposite fluctuation would
transform a DN into a NL). The behaviour of VY Scl stars
contradicts, however, this apparently reasonable conclusion. These
are very bright NLs which occasionally undergo a fall in
brightness by more than one magnitude. Although such drops in
luminosity bring them into the DN instability strip, they show no
outbursts. In fact, during their low states, which may last weeks
or several years, VY Scl stars may spectroscopically appear like
quiescent DNs but they {\sl do not} become `regular' DNs (see
Warner \cite{w95} and references therein). At best they show once
or twice a DN-type outburst but not an outburst cycle one could
expect from systems at mass-transfer rates lower than the critical
one for the disc instability to occur.

It would be tempting to say that VY Scl stars during low states
are stable with respect to the DN-type instability, i.e. the
effective temperature of their disc is below the instability
strip. Since they are also stable in their high  states, they
would just oscillate between two stable steady-state
configurations, explaining the general absence of outbursts.
However, this simple explanation cannot be correct because during
the transition from one stable state to the other, the disc
configuration (its accretion rate) must cross the dwarf-nova
instability strip and therefore show outbursts.

\subsection{Mass transfer fluctuations}

Before trying to explain this strange behaviour, one should
determine the cause of brightness fluctuations in these systems.
They seem to be due to mass-transfer rate fluctuations. Indeed,
there is compelling evidence for mass-transfer fluctuations of
various amplitudes occurring on various time-scales. Of course, in
systems with an accretion disc, it might be difficult to make the
difference between variations in the {\sl accretion} and the {\sl
mass-transfer} rates, unless one can observe the
mass-transfer-stream impact region in the outer disc (the so
called `bright' or `hot' spot) whose changes of brightness would
reflect mass-transfer variations. However, in polars (AM Her
stars), a class of strongly magnetized CVs, no accretion disc is
present so that observed changes in brightness by 1.5 do 4.5 mag
must result directly from mass-transfer variations (see
Hessman et al. \cite{hgm00} and Buat-M\'enard et al. \cite{bhl01b}
for an evaluation of the effect in DNs of mass transfer variations
similar to those observed in AM Her systems). Also in VY Scl
stars it is obvious that the mass-transfer rate fluctuates between
a very high value (comparable or higher than the {\sl accretion
rate} at DN's maximum; Warner \cite{w95}) and a value so low that
the mass-transfer rate can be considered to have stopped.

Let us first consider what is happening when the mass-transfer
rate is rapidly switched off. In such a case, as nicely explained
by Leach et al. (\cite{lhk99}, hereafter L99), the moment the
mass-transfer rate drops below the (upper) critical value the
outer disc becomes unstable and a cooling front propagates
inwards, bringing the whole disc into a cold state. This state,
however, is {\sl non-stationary} as the accretion rate in the disc
is not constant but increases with radius (see e.g. Figs. 10 and
17 in Lasota \cite{l01}). Since in such a disc most of the mass is
in its outer regions, matter will diffuse inwards and inevitably
cross the critical line as in Figs. 10 and 17 of Lasota
(\cite{l01}) and the system will go into outburst.

L99 suggested that in VY Scl irradiation by white dwarfs heated by
accretion to very high temperatures ($\gta 40,000$ K) during high
states, stabilizes the inner disc thus preventing outbursts.
Indeed, as noticed by Lasota et al. (\cite{lhh95}) a cold,
dwarf-nova-type accretion disc (i.e. a disc whose structure
results from the passage of a cooling front) can be stable if its
inner part is removed. In L99 it is assumed that the inner disc is
kept hot by white-dwarf irradiation which, as far as the
cooling-front propagation is concerned, has the same effect a disc
truncation: the cooling front has to stop at the transition radius
between the cold and the hot part of the disc. The transition
radius has to be large enough to allow the cold disc to settle to
a quasi-stationary stable configuration (see L99 for details). As
we will see in Section 2 the L99 model cannot work for white-dwarf
masses larger than $\sim 0.4$ \msol, which makes its application
to {\sl all} VY Scl stars rather questionable. In Section 3 we
will show, however, that inner disc truncation by magnetic fields
of not extravagantly large values (magnetic moments $\mu\gta
5\times 10^{30}$ G cm$^{3}$, i.e. magnetic fields $\gta 40$ kG for
a $\sim 0.7$ \msol white dwarf) also suppress outbursts in VY Scl
low states and does not require unrealistic parameters. The
proposal that VY Scl systems may contain magnetized white dwarfs
is not new: using arguments based on the observed spectral or
timing properties of these systems, Jameson et al. (\cite{jsk82}),
and Hollander \& van Paradijs (\cite{hv92}) proposed that TT Ari
might be magnetized, while Voikhanskaia \cite{v88} suggested that
accretion in MV Lyr occurs on a magnetic pole during its low
states.

In Section 4 we point out that although such a truncation
suppresses outbursts during the minimum light, it cannot
explain the absence of outbursts during the very slow decays from
maximum to minimum light and very slow rises back to maximum. For
example, Honeycutt et al. (\cite{hlr93}) observed a $\sim 500$
day, 4 mag rise of \object{DW UMa}, which is very long compared to
the disc viscous time expected to be rather 10 - 20 days. It is
instructive to have a closer look at Fig. 4 of L99. It shows part
of the \object{MV Lyr} light-curve and the light-curve
calculated by L99. Their model nicely fits the first rapid drop in
luminosity but fails to reproduce the following slow rise and had
these authors continued their calculation it would miserably fail
to account for next, very slow (50 days) decay (see our Fig.
\ref{fig:slowvar}), contrary to their expectations.

The reason is simple: when the mass-transfer rate remains in the
instability strip longer than the viscous time, the system must produce
a series of outbursts. We will show that in order to suppress outbursts
during very slow rises and decays one has to suppress the disc as soon
as the mass-transfer rate enters the instability range. This, as we
argue, requires magnetic moments similar to those of Intermediate
Polars,

In Section 5 we recall the independent evidence for the magnetic
nature of VY Scl stars. In Section 6, after discussing the
possibility that in some VY Scl stars the magnetized white dwarf
could be in the propeller regime (we modify the standard inner
boundary condition of the model in order to account for a source
of angular momentum and matter ejection), we decide this would not
change our main conclusion about the magnetic moment required to
stabilize intermediate luminosity states. In Section 7 we discuss
the relation between VY Scl stars and other types of CVs. Section
8 summarizes our results.

\section{Irradiation by a hot white dwarf}

As mentioned above, L99 proposed that in the low-state, the inner
disc of VY Scl stars is kept in a hot and stable state by
white-dwarf irradiation as in King (\cite{k97}). Because most of
the time they accrete matter at very high rates, white dwarfs in
VY Scl stars are hotter than in most dwarf novae ($T_{\rm
eff}\sim 35,000 - 65,000$ K) so the L99 proposal seems to be
reasonable. As described by these authors, the cooling front,
which propagates inward when the mass-transfer rate falls below
the critical value for instability, will reach only the boundary
with the hot region. If this region is large enough the disc left
behind the cooling wave will be cold and stable. The condition for
this to happen can be written as:

\begin{equation}
r_{\rm tr} > r_{\rm crit} \approx 6 \times 10^{9} \left(\frac{\dot
M}{10^{15}{\rm g\ s^{- 1}}}\right)^{0.375} M_1^{0.333}\; {\rm cm},
\label{eq:rin}
\end{equation}
where $r_{\rm tr}$ is the transition radius and $M_1$ is the
white-dwarf mass in solar unit (see e.g. Lasota \cite{l01}; Lasota
et al. \cite{lkc99}). The quiescent disc would thus have two
components: very hot ($T_{\rm eff}> 7000$ K) for  $r < r_{\rm
tr}$; very cold ($T_{\rm eff}< 3000$ K) for  $r > r_{\rm
tr}$.

\begin{figure}
\resizebox{\hsize}{!}{\includegraphics{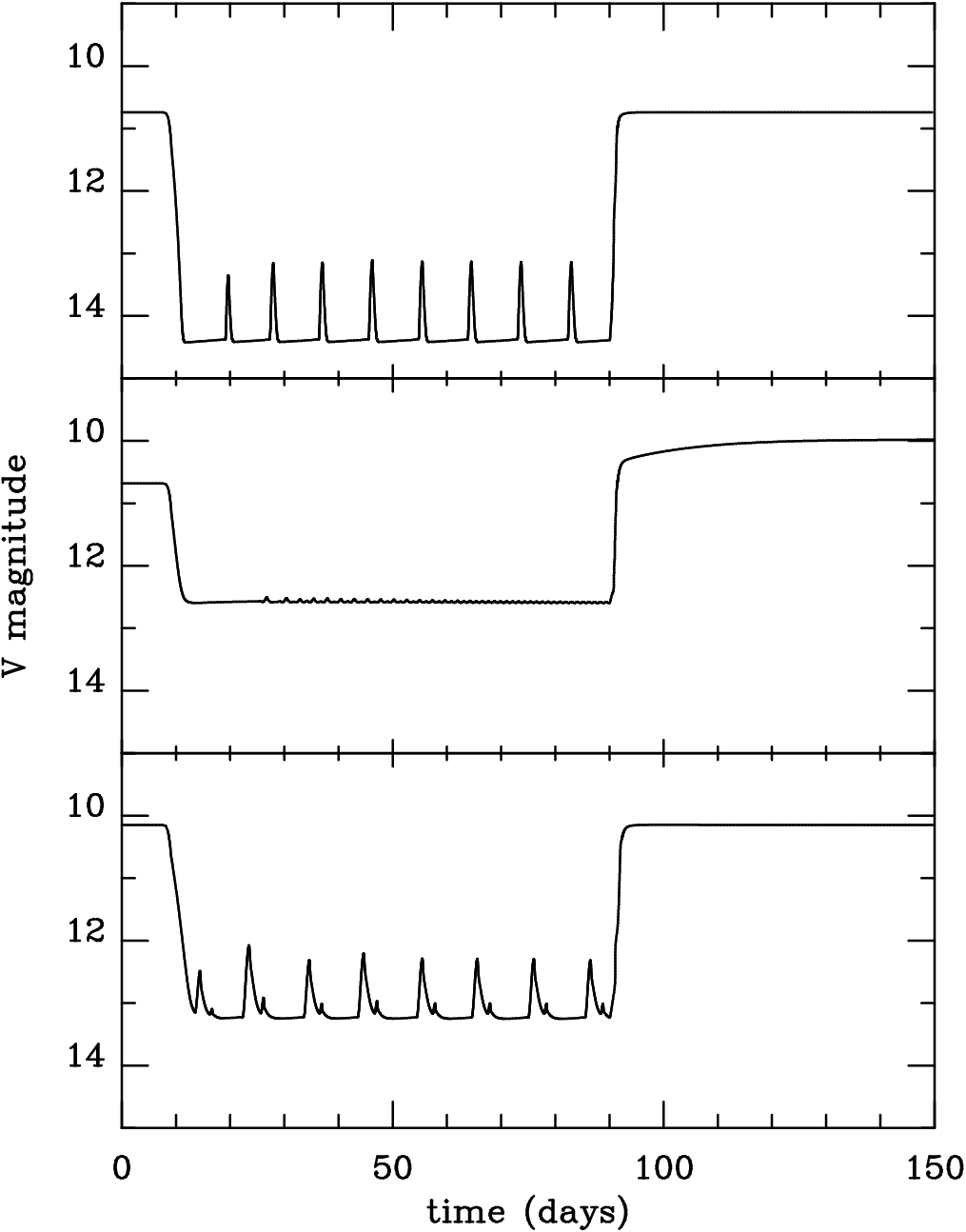}}
\caption{Visual light curves of a binary system in which the mass
transfer from the secondary stars is switched off for 80 days and
then brought back to its previous value $\dot M_{\rm tr}=- 2
\times 10^{17}$ g s$^{- 1}$. The upper and intermediate panels
correspond to the cases calculated by L99, i.e. to 0.4 \msol white
dwarf heated to $T_* = 20,000$ and 40,000 K respectively. The
lower panel corresponds to a 40,000 K, 0.7 \msol white dwarf. }
\label{fig:nmag}
\end{figure}

We repeated L99 calculations using Hameury et al. (\cite{hmd98})
code as modified by Buat-M\'enard et al. (\cite{bhl01a} -- we have
corrected Eq. (1) of this paper, in which the heating rate was
overestimated by a factor $\sim$ 2), treating disc irradiation by
the white dwarf as in Hameury et al. (\cite{hld99}, see also Dubus
et al. \cite{dlh99}). The main difference between our code and the
one used by L99 is that in their case the outer radius is kept
fixed and heating by tidal torques and stream-heating are
neglected, while in our case the outer radius is allowed to vary
(in agreement with both angular momentum conservation and
observations) and we include in our calculation external heating
of the outer disc. The difference in the heating term is not very
important in this context. We considered first the case of a 0.4
\msol white dwarf and the same disc parameters as in L99. The
high-state mass-transfer rate is $\dot M_{\rm tr}=- 2 \times
10^{17}$ gs$^{-1}$, lower than the value used in L99, since
heating of the outer disc by the tidal torques and the stream
impact lowers the critical mass transfer rate above which the disc
is stable. To trigger a low state the mass-transfer is switched
off. We considered the same two white-dwarf temperatures as L99:
20,000 and 40,000 K. The results, very similar to those of L99 are
shown in Fig. \ref{fig:nmag}. Irradiation by a 40,000 K white
dwarf suppresses dwarf-nova type outbursts, although some residual
mini-outbursts (or rather micro-outbursts) are still present.
Their amplitudes are too small to be of observational importance.
They are of physical, not numerical origin. As explained in
Hameury et al. (\cite{hld99}), the intermediate region (at
$r=r_{\rm tr}$), where the temperature ranges from 3000 to 7000 K,
is violently unstable and generates a series of `mini-outbursts'.
When $r_{\rm tr}< r_{\rm crit}$ these `mini-outbursts' can be
quite prominent (Hameury et al. \cite{hld99}, \cite{hlw00}), but
here they are rather inconspicuous.

L99 used only one value of the white dwarf mass, $M_1 \approx
0.41$ - the (apparently well determined) mass of the primary in
\object{LX Ser} - an eclipsing system. This value, however, is
crucial for the effect to work because the larger the white dwarf,
the larger the inner-disc hot region. As shown in Fig.
\ref{fig:nmag}, for a typical NL mass of 0.7 ${\rm M_{\odot}}$,
even a 40,000 K white dwarf cannot prevent a series of prominent
outbursts. We checked numerically how the outburst amplitude
decreases with increasing white dwarf temperature; for 50,000 K,
we get 0.6 mag outbursts, and for 60,000 K, the amplitude is only
0.25 mag.

This is can be also seen without using numerical codes. The white-dwarf
irradiation temperature is given by (e.g. Smak \cite{s89})
\begin{equation}
T_{\rm irr}^4 = (1-\beta) T_*^4 {1 \over \pi} [\arcsin \rho -\rho
(1-\rho^2)^{1/2} ]
\label{eq:till}
\end{equation}
where $\rho = R_*/r$, $R_*$ and $T_*$ are the white dwarf radius
and temperature. $1 - \beta$ is the fraction of the incident flux
which is absorbed in optically thick regions, thermalized and
re-emitted as photospheric radiation. Therefore one can estimate
the transition radius as
\begin{equation}
r_{\rm tr} \approx 6.7 \; T_{*,40}^{4/3} \left(\frac{T_{\rm
crit}}{6500\ K}\right)^{-4/3} R_* \ , \label{eq:rtr}
\end{equation}
where we put $\beta=0$ (no albedo, which obviously maximizes the
transition radius) and assumed for simplicity that a disc is
thermally stable if its photospheric temperature is larger
than $T_{\rm crit} \simeq 6500$ K. ($T_{*,40}= T_*/40,000 K$). A
comparison with Eq. (\ref{eq:rin}) shows immediately the advantage
of using a 0.4 \msol white dwarf, for which $R_*=1.07 \times
10^{9}$ cm. A 0.7 \msol would stabilize the disc only if heated to
more than 50,000 K. So unless {\sl all} VY Scl have very low-mass
and/or very hot white dwarfs the L99 model cannot explain the lack
of outbursts during the low states of these binary systems.

It is unlikely that the VY Scl phenomenon occurs only in systems
with very low-mass primaries. White dwarf masses are
notoriously difficult to measure (see e.g. the discussion by
Shafter \cite{s83}), and some care must be taken when considering
each mass determination individually, but it would be very
surprising that all mass measurements would be in error when
exceeding 0.4 M$_\odot$. For example, another eclipsing VY Scl
star, \object{BH Lyn} ($P_{\rm orb} =3.74)$ h) has a primary mass
more typical of its class: $M_1 \approx 0.7$ (Hoard \& Szkody
\cite{hs97}). Other VY Scl have also higher primary masses:
\object{KR Aur} -- $M_1 \simeq 0.70$ (Shafter
\cite{s83}),\object{V425 Cas} -- $M_1 \approx 0.86$ ; only
\object{V442 Oph} is reported to have $M_1 \approx 0.34$ (Ritter
\& Kolb \cite{rk98}). In any case 0.4 \msol is not the value of
white-dwarf masses expected in CVs with orbital periods between 3
and 4 hours. It is interesting to note that $M_1 \approx 0.4$ is a
value favored by the hypothesis of Greiner et al. (\cite{gtd99})
that some (all) VY Scl stars are X-ray supersoft sources, but this
hypothesis was rejected by Mauche \& Mukai (\cite{mm02}) on the
grounds that the X-ray spectra of VY Scl stars are not blackbodies
and by Patterson et al. (\cite{ptf01}) who argued that very high
white dwarfs X-ray luminosities resulting from the blackbody
hypothesis should also imply high optical fluxes which are not
detected.

One should finally note that not all white dwarfs in VY Scl stars
are very hot: Sion (\cite{es99}) quotes a temperature of 35,000 K
for \object{RX And}, too low to significantly affect the stability
of the disc. Therefore, heating by a hot white dwarf, even if
it may affect significantly the stability of the inner disc,
cannot be the universal explanation for the VY Scl behaviour.

\begin{figure}
\resizebox{\hsize}{!}{\includegraphics{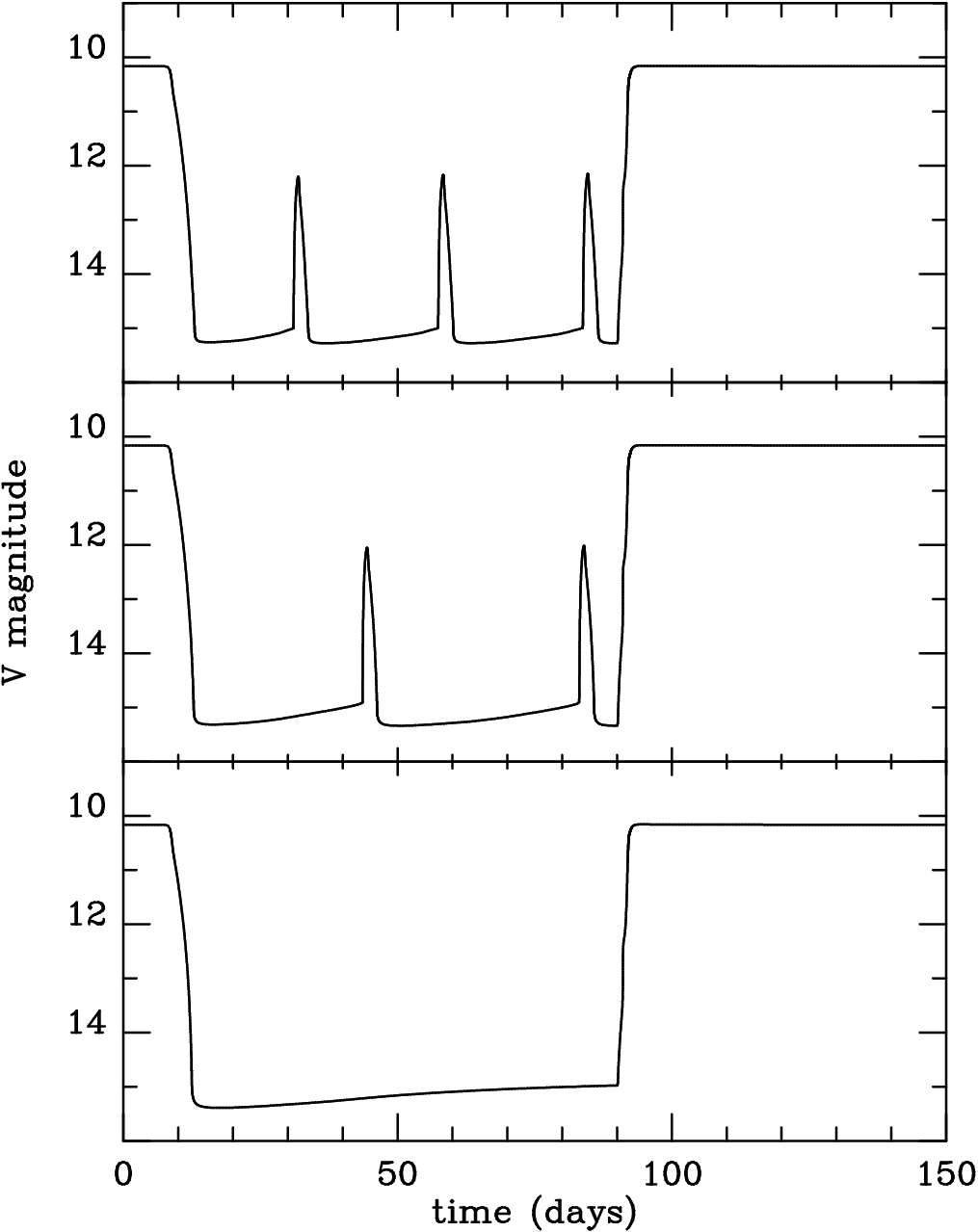}}
\caption{Visual light curves of a binary system in which the
mass-transfer rate is switched off for 80 days then switched on
back to its previous value. The accretion disc is disrupted by the
magnetic field of a 0.7 \msol white dwarf. Three values of the
magnetic moment were used: $\mu_{30}$= 1 (top panel), 2
(intermediate panel), and 5 (bottom panel).} \label{fig:mag}
\end{figure}

\section{Truncation by a magnetosphere}

We conclude that heating of the inner disc by hot white dwarfs is
unlikely, at least in general, to prevent outbursts during VY Scl-star
low states. There is, however, a different way of removing the inner,
unstable part of an accretion disc: disruption by a magnetized white
dwarf. We have already calculated dwarf-nova outburst cycles for
truncated accretion discs in Hameury et al. (\cite{hld99}) and in
Hameury et al. (\cite{hlw00}). Here, we apply the same scheme to VY Scl
stars.

The inner disc radius is given by the "magnetospheric
radius":
\begin{equation}
r_{\rm in}= r_{\rm M}= 9.8 \times 10^8 \dot M_{15}^{-2/7}M_1^{-1/7}
             \mu^{4/7}_{30} \; \rm cm
\label{eq:rm}
\end{equation}
where $\mu_{30}$ is the magnetic moment in units of $10^{30}$
Gcm$^3$. As an example we calculated the case of a binary system
with $M_1=0.7$, $P_{\rm orb}= 3.5$ h, secondary mass $M_2=0.3$ for
three values of the magnetic moment: with $\mu_{30}$= 1, 2, and 5.
The mass transfer rate $\dot M_{\rm tr}$ = 2 10$^{17}$ gs$^{-1}$
was switched off instantaneously on day 8 and switched on back 80
days later. Figure \ref{fig:mag} shows that a rather
moderate value of the magnetic moment, $\mu=5\times 10^{30}$
Gcm$^3$ is sufficient to prevent dwarf-nova-type outbursts. For a
0.7 \msol white dwarf, this value of the magnetic moment
corresponds to a magnetic field of 40 kG. Thus, even a rather weak
magnetic field would explain low-state properties of VY Scl stars.

\section{Absence of outbursts during intermediate states}

\subsection{Disrupted discs}

\begin{figure}
\resizebox{\hsize}{!}{\includegraphics{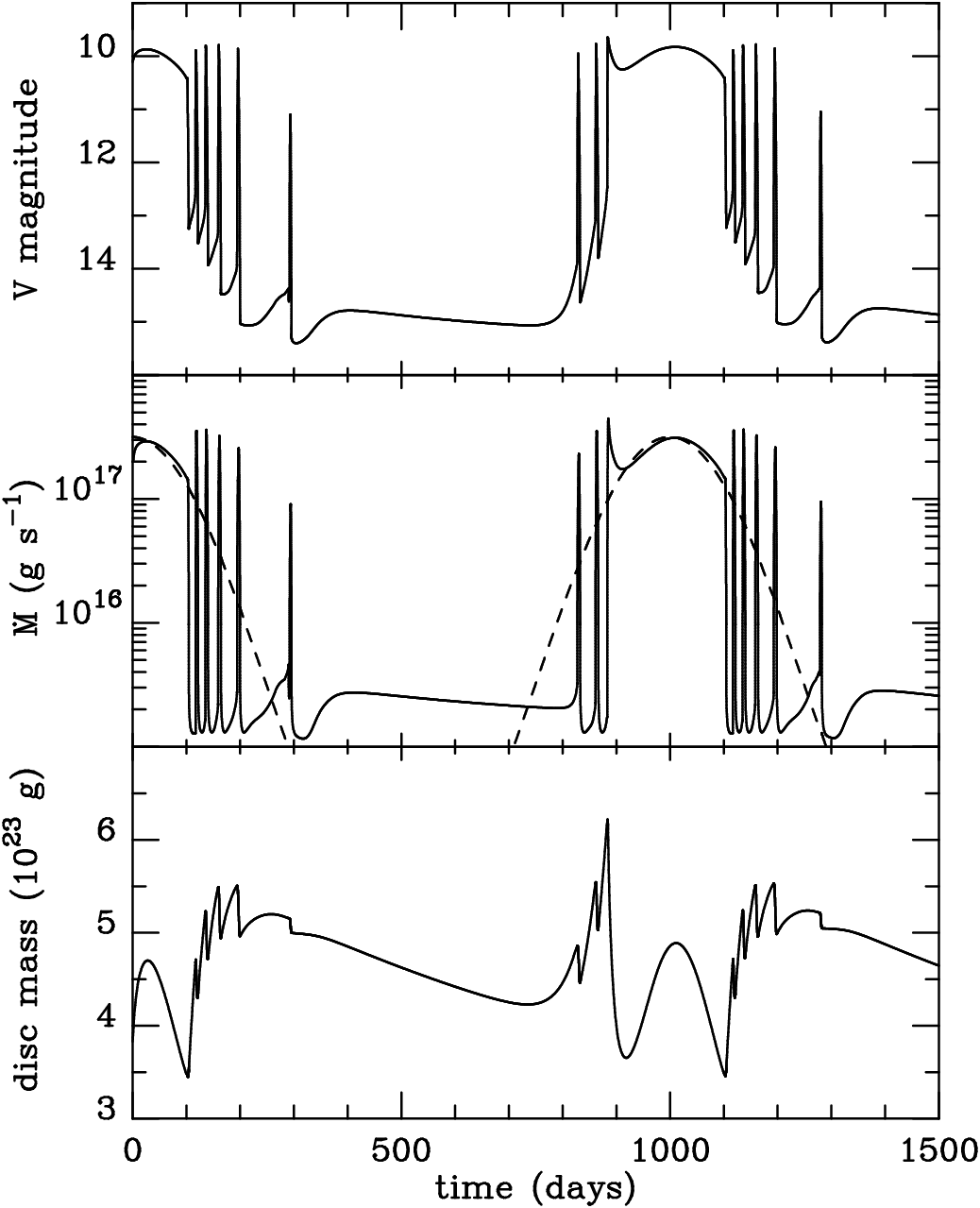}} \caption{Top
panel : visual light curve of a binary system in which the
mass-transfer rate slowly varies. The accretion disc is disrupted
by the magnetic field of a 0.7 \msol white dwarf with $\mu_{30}$=
10. Intermediate panel : mass accretion rate onto the white dwarf
(solid line), and mass transfer rate (dashed line). Bottom panel :
disc mass } \label{fig:slowvar}
\end{figure}

As mentioned in the Introduction the disc truncation sufficient to
suppress outbursts in quiescence still allows for their appearance
during intermediate states lasting longer than the disc's
characteristic viscous time (see Fig. \ref{fig:slowvar}). More
precisely the condition for the absence of outbursts in the low
states is a condition for the absence of {\sl inside-out}
outbursts, i.e. outbursts starting in the inner regions of the
disc. Inside-out outbursts occur when the viscous-diffusion time
is shorter that accumulation time at the disc's outer rim (see
Lasota \cite{l01} for a detailed discussion). When one switches
off the mass-transfer rate, the accumulation time is
obviously infinite and only inside-out outbursts can occur. That
is why inner-disc truncation suppresses outbursts in this case.
However, if the mass-transfer rate is lowered slowly from a stable
to an unstable range, the disc will enter first the regime
of {\sl outside-in} outbursts as such outbursts occur at the
highest range of unstable transfer rates, because it is
where the accumulation time $\sim 1/\dot M_{\rm tr}$ is the
shortest ("outside-in" does not mean necessarily that outbursts
start exactly at the outer edge - see e.g. Buat-M\'enard et al.
\cite{bhl01a}). During this phase, the disc slowly builds up; its
mass increases significantly above the minimum values reached
after the end of the stable hot phase, and outbursts appear in the
outer disc regions, even for values of the magnetic field for
which the disc was found to be stable when the mass transfer rate
was very rapidly shut off.

As an example, we show in Fig. \ref{fig:slowvar} what is happening
in a system with the same parameters as those assumed for
Fig.~\ref{fig:mag} but with a slow (300 days) drop (and rise) of
the mass-transfer rate. A series of outbursts appear before the
system settles to the low state in which, as before, outbursts are
absent; outbursts also reappear on the way back to maximum
brightness. (In this case the low state was found to be stable
only for magnetic moments $\mu_{30}$ = 10 because of a different
disc evolution prior to the low state).

If the VY Scl phenomenon were to be explained by irradiation
of the disc, this would require illumination to be large enough
for the outer edge of the disc to be brought to a temperature
larger than about 6500 K, i.e. $r_{\rm tr} > r_{\rm out}$. This
condition is much more stringent than the conditions for having no
outburst in quiescence which implies a much smaller $r_{\rm tr}$.
For typical disc radii of order of 2 -- 2.5 10$^{10}$ cm, this
would require white dwarf temperatures in excess of 100,000 K,
which have never been reported. Moreover, at such large
temperatures, the luminosity would be dominated by the intrinsic
and reprocessed white dwarf luminosity, the accretion luminosity
would be comparatively small, and low states very shallow.

Therefore, irradiation can perhaps explain the nature of systems
with low mass white dwarfs, in which sharp transitions are
observed between low and high states; it cannot possibly explain
the systems which show slow variations between outburst and
quiescence. As long as there is a full-fledged disc there will be
outbursts. Therefore the only remedy against outbursts is the
suppression of the disc as soon as the mass-transfer rate from the
secondary enters the unstable range, $\dot M_{\rm tr} \leq \dot
M_{\rm crit}^{\rm h}$, where $\dot M_{\rm crit}^{\rm h}$ is the
minimum accretion rate for stable, hot equilibria.

Observations tend to support this conclusion : G\"ansicke et
al. (\cite{gsb99}) found that the spectra of TT Ari during a low
state shows virtually no sign of an accretion disc. The UV and IR
continuum emission was clearly dominated by the two stellar
components, very weak lines were visible in the continuum, which
lead these authors to conclude that there was no optically thick
disk up to at least 12 white dwarf radii. Similarly, Knigge at al.
(\cite{klh00}) found that they were able to see the hot, bare
white dwarf during quiescence, which was hidden by the disc during
high states.

The only possibility we can think of for suppressing the disc is
the existence of a relatively large magnetic field; the precise
form of the required ``no-disc" condition then depends essentially
on the inner disc boundary condition.

In general this condition states that the magnetospheric
radius must be larger than some characteristic disc radius,
i.e.
\begin{equation}
r_{\rm M} > f a
\label{eq:cond1}
\end{equation}
where the "characteristic radius" is a fraction $f$ of the orbital
separation
\begin{equation}
a=3.53 \times 10^{10} (M_1 + M_2)^{1/3} P_{\rm h}^{2/3} \; {\rm
cm} \label{eq:a}
\end{equation}
$P_{\rm h}$ being the orbital period in hours. The critical
accretion rates is (see e.g. Lasota \cite{l01})
\begin{equation}
\dot{M}_{\rm crit}^{\rm hot} = 9.5 ~ 10^{15} ~ \alpha^{0.01}
M_1^{-0.89} \left( {r \over 10^{10} \rm cm} \right)^{2.68} ~\rm
g~s^{-1}, \label{eq:mdmin}
\end{equation}

As a ``no-disc" condition we can require, for example, that when
the mass transfer rate from the secondary drops below the critical
value corresponding to the end of hot branch at the outer disc's
edge, $\dot{M}_{\rm crit}(r_{\rm out})$ ($r_{\rm out}$ being the
outer disc radius), the magnetospheric radius becomes larger than
the circularization radius. From Eqs. (\ref{eq:rm}),
(\ref{eq:cond1}), (\ref{eq:a}), (\ref{eq:mdmin}) one obtains the
following general ``no-disc" condition in terms of the required
magnetic moment:
\begin{equation} \mu \gta 1.5
\times 10^{33} f^{1.75}_{0.12}P^{2.06}_{4} (3r_{\rm out}/a)^{1.34}
M_1^{1.4} {\; \rm G\ cm^3}
\label{eq:cond2}
\end{equation}
where we assumed a mass ratio $M_2/M_1=0.43$, used the value
$f=0.12$ corresponding to the circularization radius,
measured the orbital periods in units of 4 hours, and
normalized $r_{\rm out}$ by its `typical' value $a/3$.

Numerical simulations show that this is an overestimate, the critical
magnetic moment being about half the value deduced from Eq.
(\ref{eq:cond2}). This is due to the fact that heating by the tidal
torque and stream impact has a stabilizing effect in the outer disc
regions, so that Eq. (\ref{eq:mdmin}) gives an overestimate of
$\dot{M}_{\rm crit}^{\rm hot}$. Also $r_{\rm out}$ decreases
significantly when the inner disc radius approaches $r_{\rm circ}$.
So for a 0.7 \msol white dwarf in a 4 hours binary the required magnetic
field would be  $B\gta$ 6 MG .

The values of magnetic moments required to prevent outbursts
during intermediate states of VY Scl stars are in the range
between DQ Her and Intermediate Polars (see e.g. Warner
\cite{w95}). We should therefore consider what might be happening
if a VY Scl star were in the propeller regime which apparently
characterizes the DQ Her star \object{AE Aqr} (Wynn et al.
1997).

\subsection{Dwarf-nova outbursts in the magnetic propeller regime}

In deriving condition Eq. (\ref{eq:rm}) we have assumed that the
inner edge of the accretion disc corresponds to the (variable)
magnetospheric radius. In the numerical calculation the inner
boundary condition was the usual ``no--stress" condition which in
practice corresponds to the condition $\nu \Sigma|_{\rm in}=0$. Of
course this condition is not the right one in the propeller regime
and in general when there is an angular momentum source at the
inner disc edge. Livio \& Pringle (\cite{lp92}) modified the
surface-density diffusion equation to include the existence of a
torque due to the interaction of the white dwarf magnetic field
with the accretion disc, keeping the inner disc radius equal to
that of the white dwarf. Duschl \& Tscharnuter (\cite{dt91}) on
the other hand considered the case of an extended boundary layer,
which also changed the $\nu \Sigma = 0$ inner boundary condition,
with a disc inner radius larger than the white dwarf radius. Here
we find a boundary condition which is appropriate for the magnetic
propeller regime.

The magnetic torque $T$ exerted on a small portion $\Delta r$ of
the disc can be written as (Livio \& Pringle \cite{lp92}):
\begin{equation}
T = {B_z^2 \over 4 \pi} {\Omega_{\rm in} - \Omega_* \over
\Omega_{\rm in}} 4 \pi r_{\rm in}^2 \Delta r
\end{equation}
where $r_{\rm in}$ is the inner disc radius, $B_z$ the vertical
component of the magnetic field, which can be assumed to vary as
$r^{-3}$, and $\Omega_{\rm in}$ and $\Omega_*$ are the Keplerian
and stellar angular velocities respectively. The inner radius should
be equal to the magnetospheric radius, defined by:
\begin{equation}
{B_z^2(r_{\rm in}) \over 8 \pi} = {\dot{M}\over 4 \pi r_{\rm in}^2} \Omega_{\rm
in} r_{\rm in}
\end{equation}
which gives:
\begin{equation}
T = 2 \dot{M} j {\Omega_{\rm in} - \Omega_*\over \Omega_{\rm in}}
\end{equation}
for $\Delta r \sim r$, $j$ being the Keplerian specific angular
momentum. This torque has to be balanced by the viscous torque at
the inner disc edge, $T_{\rm visc} = 3 \pi \nu \Sigma j$,
leading to the new boundary condition at $r = r_{\rm in}$ :
\begin{equation}
\nu \Sigma = {\dot{M} \over 3 \pi} f \left| {\Omega_{\rm in} -
\Omega_* \over \Omega_{\rm in}} \right|
\end{equation}
which is similar to that obtained by Duschl \& Tscharnuter
(\cite{dt91}) in a slightly different context; $f \leq 1$ here
accounts for the fact that matter will be ejected from the system,
carrying a fraction of the flow of angular momentum deposited in
the disc.

The steady state solution is now:
\begin{equation}
\nu \Sigma = {\dot{M} \over 3 \pi} \left[ 1 + \left(  {f
|\Omega_{\rm in} - \Omega_* |\over \Omega_{\rm in}}  -1 \right)
\left({r_{\rm in} \over r}\right)^{1/2}\right]
\label{eq:diss}
\end{equation}

Because $r_{\rm in}$ need not be much smaller than the outer disc
radius, the whole disc could be affected by this new boundary
condition when $\Omega_{\rm in} \ll \Omega_*$, i.e. when the
propeller effect is efficient. The surface density is increased as
compared to the standard case; the increase in luminosity (the
dissipation, proportional to $\nu \Sigma$, is due to the
work done by the torque, the energy source being the rotation of
the white dwarf which is partly redistributed over the whole
disc).

The local thermal equilibrium equations are unchanged, so that the
S-curves are unaffected by the propeller effect, which therefore
lowers the minimum mass transfer for the disc to remain on the hot
branch, and thus has a stabilizing effect. The change in the
critical $\dot{M}$, however, remains moderate: for $r_{\rm
out}/r_{\rm in}=6$ and $\Omega_* / \Omega_{\rm in}=5$, Eq.
(\ref{eq:diss}) predicts a change by a factor 3.7 in $\dot{M}$,
which would in principle lower the critical magnetic moment for
which the disc would disappear before becoming unstable (see
section 4). However, the ``no-disc" criterion is also changed; a
disc can still exist even for a circularization radius larger than
the inner radius, because of the torque exerted by the
magnetosphere. Both effects compensate; numerical calculations
showed that Eq. \ref{eq:cond2} is still a reasonably good
approximation of the critical magnetic moment.

Finally, it should be noted that, even in the propeller
regime, pulsations at the white dwarf spin period are expected. AE
Aqr for example clearly show these pulsations which can be due
either to residual accretion or to magnetic dissipation at the
poles (Wynn et al. \cite{wkh97}).

\section{VY Scl stars as intermediate polars ?}

Eq. (\ref{eq:cond2}) states that (at least some) VY Scl stars have
magnetic fields close to those of intermediate polars (IPs). This
conclusion follows solely from photometric properties of VY Scl
stars during their intermediate states as confronted with
requirements of the disc instability model. Obviously, to be
treated seriously, it better be supported by some direct evidence
such as circular polarization and/or pulsations reflecting the
presence of a spinning magnetized white dwarf. No such evidence
has been found yet.

However, three (\object{LX Ser}, \object{BH Lyn} and \object{DW
UMa}) VY Scl stars (see e.g. Hellier \cite{h00}) are classified at
high state as SW Sex stars (to which one could add \object{PG
And}, a SW Sex not classified as VY by Hellier \cite{h00}, but
which has occasional low states, Thorstensen et al.
\cite{trw91}). These are nova-like CVs showing anomalous
phenomena in their emission lines, where ``anomalous" means
different from the simple binary model of CVs (see e.g. Warner
\cite{w95}). Also their continua show unusual temperature profiles
(see however Smak \cite{s94} for a discussion of the bias
introduced by the high inclination on the determination of the
temperature profile). These anomalies are detected in systems with
high inclinations, but seem in fact to be shared by most CVs at
high accretion rates (Horne \cite{h99}).

In two SW Sex stars, \object{LS Peg} (Rodr\'{\i}guez-Gil et al.
\cite{rcm01}) and \object{V795 Her} (Rodr\'{\i}guez-Gil et al.
\cite{rcm02}), variable circular polarization has been discovered.
The authors of these observations deduce magnetic fields of 5--15
MG (\object{LS Peg}) and 2--7 MG (\object{V795 Her}). It is
interesting to notice that \object{V795 Her} has been an IP
candidate for a long time (it is listed as an ``DQ Her" star in
the Downes et al. (\cite{dws01}) catalogue and atlas of
cataclysmic variables). The circular polarization in \object{LS
Peg} and \object{V795 Her} is modulated with periods of $\sim 20$
-- 30 min, typical of IPs and fitting the $P_{\rm spin}/P_{\rm
orb} \sim 0.1$ relation (see e.g. King \& Lasota \cite{kl91}).
These observations provide indirect support to our conclusion
about the magnetic nature of VY Scl stars.

Groot et al. (\cite{grp01}) having observed \object{SW Sex} itself
-- in an ``excited, low state", which probably means fainter
than maximum by less than about 1 mag, but no value of $m_V$ is
given in their paper -- find a disc structure which could be
compatible with a IP-like white dwarf. These authors however were
sceptical about the magnetic nature of SW Sex stars in general
because, apparently unaware of the results of Rodr\'{\i}guez-Gil
et al. (\cite{rcm01}, \cite{rcm02}), they worried about the
absence of pulsations and polarization.

Our results could imply that variable circular polarization should
be detected, at least in such VY Scl stars as \object{DW UMa} or
\object{MV Lyr} in which long intermediate states are observed.
One should keep in mind, however, that circular polarization has
been detected only in five IPs out of a total of more than 30, the
ones that harbor white dwarfs with highest magnetic fields $\sim 2
- 8 MG$ (see Warner \cite{w95}). Since VY Scl stars could have
fields lower than this value the detection of their circular
polarization could be rather difficult. Indeed, Tapia (1981,
quoted in Robinson et al. \cite{rbc81}) found the circular
polarization of \object{MV Lyr} to be less than 0.13\% and
consistent with zero. This should be compared with the circular
polarization of \object{V795 Her} - mean level increasing from
0.15\% in $U$ to 0.41\% in $I$ - and the 0.3\% amplitude in
\object{LS Peg} (Rodr\'{\i}guez-Gil et al. \cite{rcm01},
\cite{rcm02}).

We do not know of VY Scl pulsation periods being reported in the
literature, except for a `negative superhump' (i.e. a strong
periodic signal displaced from the orbital period by few percent)
in \object{V751 Cyg} (Patterson et al. \cite{ptf01}), which could
reflect the magnetic white-dwarf's rotation.

\section{Mass transfer variations : relation to other systems}

{Fig. \ref{fig:pdist} shows the distribution of subclasses of CVs.
Here, the data is from the Ritter \& Kolb \cite{rk98} catalog,
except for magnetic systems, which are often detected by their
X-ray emission, and whose number has raised significantly in the
last few years; these are taken from the online version of the
Downes et al. (\cite{dws01}) catalog. We also used the recently
found orbital periods of \object{V794 Aql} - Honeycutt \&
Robertson \cite{hr98} and \object{V751 Cyg} - Patterson et al.
\cite{ptf01} and we do not consider \object{RX And}, $P_{\rm orb}=
5.04$ h, to be a VY Scl star, see below).} {\sl All} VY Scl stars
have orbital periods between 3.19 and 3.99 h -- the strongest
correlation observed among CVs (see Warner \cite{w00}. It is
interesting to compare VY Scl stars with the other class of
systems showing large mass-transfer fluctuations: the AM Her stars
and IPs.

\subsection{Magnetic systems}

Only six AM Her stars are known to have periods longer than 4
hours and their long-term light curves do not seem to be known.
\object{AM Her}, the prototype and best studied polar, has an
orbital period of 3.09 h and exhibits high and very low states
(see Hessman  et al. \cite{hgm00}). Another type of CVs present in
the 3 -- 4 hour interval of orbital periods are IPs, already
mentioned above. Only two of these systems showed well-documented
very low states: \object{V1223 Sgr} (which has a light curve
similar to that of \object{AM Her}; Warner \cite{w95}) and
\object{AO Psc}, with orbital periods of 3.366 h and 3.591 h
respectively. On the other hand, \object{FO Aqr} has shown no
distinct low state since 1923 (Warner 1995) and its orbital period
is equal to 4.849 h. It is therefore plausible that the
orbital-period interval between 3 and 4 hours is characterized by
large amplitude mass-transfer fluctuations. The lower limit of
this interval (3 h) marks the upper edge of the famous ``period
gap" whose lower edge is around 2 hours. One should notice,
however, that the mean mass-transfer rate in AM Her stars and IPs
is lower than in VY Scl and other NLs of \object{SW Sex},
\object{UX UMa} and \object{RW Tri} type. $\dot M^{IP}_{\rm
transf} \lta 2 \times 10^{17}$ g s$^{-1}$, while $\dot M^{VY}_{\rm
transf} \sim 2 - 8 \times 10^{17}$ gs$^{-1}$, corresponding to a
mean visual magnitude brighter than Z Cam standstills and
sometimes even brighter than dwarf-nova maxima.

\begin{figure}
\resizebox{\columnwidth}{!}{
\includegraphics{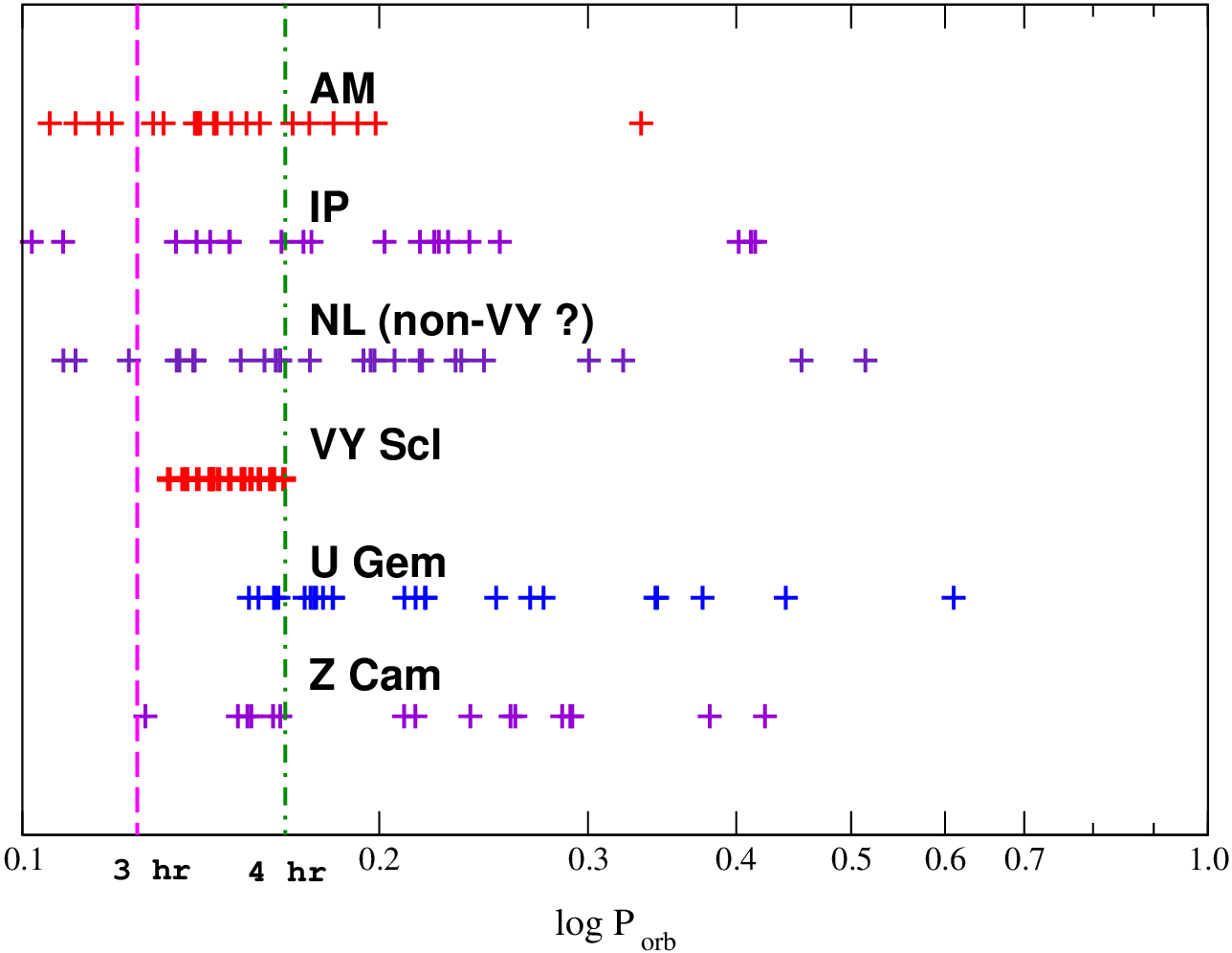}}
\caption{Orbital period distribution of various classes of CVs
between 2.4 and 15 hours} \label{fig:pdist}
\end{figure}

As seen in  Fig. \ref{fig:pdist} not all systems between 3 and 4
hours are VY Scl stars, IPs or polars. Of course, since low states
of VY Scl stars are rather rare, it is still possible that some,
or all, nova-like systems are not yet detected as VY Scl stars.
This would then indicate that many systems in the 3-4 hr
period interval contain a magnetic white dwarf (of order of
10$^{30}$ G), some of them (those showing slow declines/rises)
having field strength typical of IPs. Interestingly, quiescent
novae also seem to be magnetic when their periods are in this
range (Warner \cite{w02}). The reason for this is not clear, and
could be related to some selection effect. There is nevertheless
no reason to assume that the distribution of magnetic fields is
the same for isolated white dwarfs and for accreting white dwarfs
in a binary.

\subsection{Dwarf novae}

As for the other systems, Shafter (1992) noticed that there is a
dearth of dwarf novae in the orbital period range between $\sim 3$
and $\sim 4$ h. Indeed, according to latest edition of the CV
catalogue (Ritter \& Kolb \cite{rk98}) the shortest orbital period
for a {\sl bona fide}, i.e. U Gem-type dwarf nova is 3.8 hrs. All
other confirmed DNs (6 systems) with secure orbital periods are Z
Cam stars for shorter periods.

Mass-transfer fluctuations are also observed in DNs, both
during outbursts (hot spot brightening, see e.g. Smak \cite{s95})
and quiescence (`low' states; see Buat-M\'enard et al.
\cite{bhl01b} for references). However, Buat-M\'enard et al.
(\cite{bhl01b}) calculated, in the framework of DIM, the impact
mass-transfer fluctuation would have on the DN outburst cycle and
concluded (contrary to Schreiber et al. \cite{sgh00}) that in
these systems their amplitude must be orders of magnitude lower
than in polars.

Buat-M\'enard et al. (\cite{bhl01b}) also showed that the Z Cam
``standstill'' phenomenon, i.e. the settling of luminosity during
decline from outburst at $\sim 0.7$ mag lower than the maximum for
several months or even years is very well reproduced by allowing the
mass-transfer rate to vary by about 30\% about the critical value. It
could be that many (all?) U Gem stars are as yet unrecognized Z Cam
stars (Warner \cite{w95}; Buat-M\'enard et al. \cite{bhl01b}).

\begin{figure}
\resizebox{\columnwidth}{!}{
\includegraphics{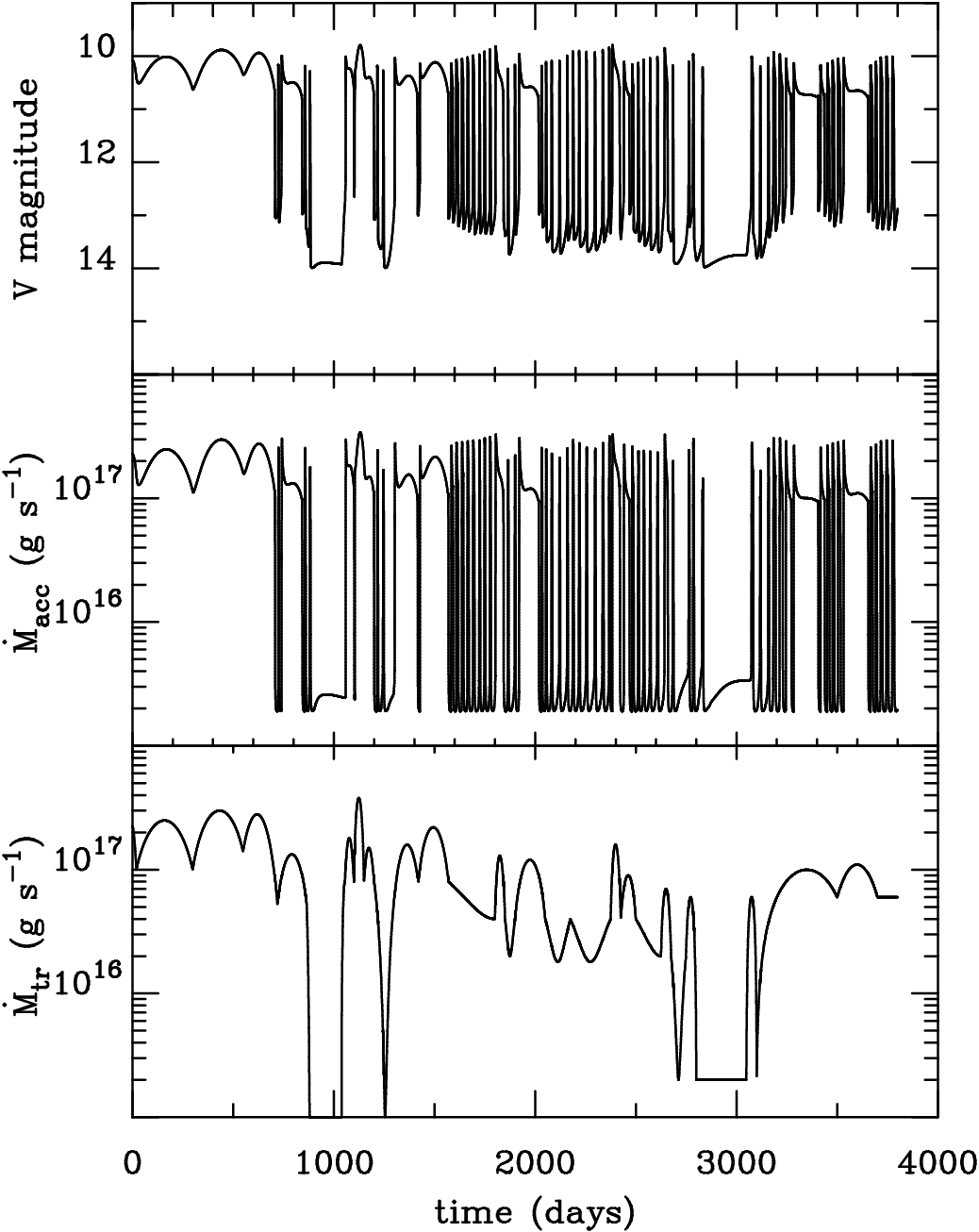}}
\caption{Variability of a hypothetical system in which the
time-dependent mass-transfer rate in AM Her (Hessman et al.
\cite{hgm00}) has been multiplied by 10 to bring it into the VY
Scl regime. Top panel: visual magnitude; intermediate panel: mass
accretion rate; bottom panel: mass transfer rate from the
secondary.} \label{fig:rxand}
\end{figure}

\subsection{From small to large $\dot{M}$ variations : the case of
\object{RX And}}

A fascinating test case is provided by \object{RX And}, the
platypus of CVs: half Z Cam, half a VY Scl star (Schreiber
et al. \cite{sgm02}). Since it is a dwarf nova during long
stretches of its life, the magnetic moment of its white
dwarf cannot satisfy Eq. \ref{eq:cond2} as a disc is
needed to produce outbursts. This is confirmed by the absence of
long intermediate state with no outbursts: VY Scl-type brightness
drops and rises are fast and when the system settles to a long
intermediate state below the standstill luminosity it shows an
outburst activity. The standstill itself corresponds to a stable
mass-transfer rate (Buat-M\'enard et al. \cite{bhl01b}). A system
exhibiting both dwarf-nova outbursts and long and quiet luminosity
descents below the stability limit would contradict our model and
presumably the disc instability model in general.

Figure \ref{fig:rxand} shows a simulation of a system in which the
mass transfer rate is 10 times that determined by Hessman et al.
(\cite{hgm00}) for \object{AM Her}; the white dwarf mass is 0.7
M$_\odot$, its temperature 40,000 K, and its magnetic moment 2
10$^{31}$ G cm$^{-3}$. The system changes from a VY Scl type to a
Z Cam one; it must however be noted that, contrary to the case of
RX And, the standstills are not always 0.5 - 1 mag fainter than
the outburst peak, simply because the mass transfer rate is not
close to the stability limit. In addition, one can see relatively
large amplitude fluctuations during the standstills, and more
generally, one expects these fluctuations during the high state of
VY Scl. This probably means that there must be a self-regulating
mechanism that maintains the mass transfer rate approximately
constant during the high state, possibly as a result of
illumination of the secondary. In any case AM Her-type variability
cannot be universal, contrary to the assertions of Hessman et al.
(\cite{hgm00}); in particular, the amplitude of the mass transfer
fluctuations in Z Cam systems must be small (Buat-M\'enard et al.
\cite{bhl01b}) .

It seems therefore that between 4 to 3 hours, dwarf novae -
systems with mass-transfer rates in the instability strip - are
(gradually?) replaced by systems in which the mass-transfer rates
is higher and whose fluctuation have also larger amplitudes.
\object{RX And} is probably a system undergoing such a transition
(Schreiber et al. \cite{sgm02}).In other words, with increasing
orbital period both mass-transfer rate and the amplitude of its
fluctuations would decrease. Large amplitude fluctuations would
die out above the orbital period of 4 hours. The decrease of
mass-transfer rate with period is required by the recent work
comparing binary evolution with observations (Baraffe \& Kolb
\cite{bk00}).

The precursors of VY Scl stars, i.e. systems with orbital periods
larger than 4 hours could be IPs or dwarf-novae, or  a combination
of the two, as \object{TV Col} and \object{V1223 Sgr} (see Warner
\cite{w95} and references therein; the orbital period of the
second system is 3.366 h).

The origin of fluctuations and their relation with the vicinity of
the period gap has been awaiting explanation since time when
Robinson et al. (\cite{rbc81}) suggested that low states of
\object{MV Lyr} could be attempts to enter the period gap and
remarked that deep low states appear when the mass-losing star
becomes fully convective.

\section{Discussion and conclusions}

We found that irradiation by a hot white dwarf cannot account for
the whole VY Scl syndrome. Even if these systems had the required
(low) white dwarf masses and very high white dwarf temperature to
prevent dwarf-nova type outbursts during their low states, the
absence of outbursts during long intermediate states observed in
some VY Scl stars cannot be accounted for by white dwarf
irradiation. Therefore in such binaries the white dwarf has to
possess a magnetic moment similar to that of IPs. This would also
prevent outbursts in the low states. On the other hand in VY Scl
systems in which fluctuations of the mass-transfer rate happen on
timescales shorter than the disc's viscous time and/or outbursts
occur during the long transitions, the stabilizing effect could be
produced by a much weaker magnetic field.

Of course we have not {\sl proven} that VY Scl stars must be
magnetic. In principle one could imagine an ``evaporation
mechanism" so tuned that it would get rid of the disc in the right
moment. Our model, however, provides the most conservative
solution of the VY Sc puzzle. In addition there exist independent
indication of the magnetic nature of least some bright nova-like
binaries.

Timing and polarization observations of VY Scl stars will provide the
ultimate test of our model.

\section*{Acknowledgments}

This work was supported in part by a grant from the {\sl Programme
National de Physique Stellaire} of the CNRS. We are grateful to
Guillaume Dubus, Andrew King and Kristen Menou for interesting
discussions, suggestions and comments. We thank the anonymous for
his careful reading of our paper and comments which helped to
improve the clarity of this paper.

\listofobjects

\end{document}